\documentclass[10pt,twocolumn,letterpaper]{article}

\usepackage[pagenumbers]{cvpr} 

\usepackage[dvipsnames]{xcolor}

\definecolor{cvprblue}{rgb}{0.21,0.49,0.74}
\usepackage[pagebackref,breaklinks,colorlinks,citecolor=cvprblue]{hyperref}

\def\paperID{*****} 
\def\confName{CVPR}
\def\confYear{2024}

\title{Efficient Feature Extraction and Late Fusion Strategy for Audiovisual Emotional Mimicry Intensity Estimation}

\author{
Jun Yu$^1$, Wangyuan Zhu$^1$\thanks{Corresponding author}, Jichao Zhu $^1$ \\
$^1$University of Science and Technology of China\\
\tt\small \{harryjun\}@ustc.edu.cn\\
\tt\small \{zhuwangyuan, jichaozhu\}@mail.ustc.edu.cn
}

\begin{document}
\maketitle
\begin{abstract}
In this paper, we present the solution to the Emotional Mimicry Intensity (EMI) Estimation challenge, which is part of 6th Affective Behavior Analysis in-the-wild (ABAW) Competition.The EMI Estimation challenge task aims to evaluate the emotional intensity of seed videos by assessing them from a set of predefined emotion categories (i.e., "Admiration", "Amusement", "Determination", "Empathic Pain", "Excitement" and "Joy"). To tackle this challenge, we extracted rich dual-channel visual features based on ResNet18 and AUs for the video modality and effective single-channel features based on Wav2Vec2.0 for the audio modality.  This allowed us to obtain comprehensive emotional features for the audiovisual modality.  Additionally, leveraging a late fusion strategy, we averaged the predictions of the visual and acoustic models, resulting in a more accurate estimation of audiovisual emotional mimicry intensity. Experimental results validate the effectiveness of our approach, with the average Pearson's correlation Coefficient($\rho$) across the 6 emotion dimensionson the validation set achieving 0.3288.

\end{abstract}    
\section{Introduction}
\label{sec:intro}

In recent years, the study of emotional mimicry intensity estimation has gained significant attention in the fields of affective computing and human-computer interaction. Emotional mimicry refers to the phenomenon where individuals unconsciously imitate the emotional expressions of others, which plays a crucial role in social interaction and empathy\cite{ding2023speed,ding2022letr}. The ability to accurately estimate the intensity of emotional mimicry from audiovisual cues is essential for developing intelligent systems capable of understanding and responding to human emotions effectively\cite{he2022multimodal}.In this paper, we present our comprehensive solution to the EMI Estimation challenge of 6th Affective Behavior Analysis in-the-wild (ABAW) Workshop and Competition\cite{kollias20246th}.

The EMI Estimation challenge\cite{kollias20246th} task aims to investigate emotional mimics through the introduction of a novel and extensive dataset. For this challenge, participants are tasked with employing a multi-output regression approach to
predict the intensities of six self-reported emotions: Admiration, Amusement, Determination, Empathic Pain, Excitement and Joy. These emotions are specifically
related to decision-making in emotional categories.

In this paper, we propose an efficient feature extraction approach combined with a late fusion strategy for audiovisual emotional mimicry intensity estimation. Our method leverages both auditory and visual modalities to capture comprehensive information about emotional expressions. By extracting discriminative features from audio and video signals, we aim to effectively represent the complex dynamics of emotional mimicry. Furthermore, we introduce a late fusion strategy to integrate the information from both modalities at a later stage of the processing pipeline. 

In general, the contributions of our work are as follows:
\begin{itemize}
    \item we extracted rich dual-channel visual features (ResNet18, AUs) and effective single-channel acoustic features (Wav2Vec2.0). This allowed us to obtain comprehensive emotional features for the audiovisual modality.
    \item leveraging a late fusion strategy, we averaged the predictions of the visual and acoustic models, resulting in a more accurate estimation of audiovisual emotional mimicry intensity. 
    \item Experimental results validate the effectiveness of our approach, with the average $\rho$ across the 6 emotion dimensionson the validation set achieving 0.3288.
\end{itemize}

The remaining structure of this paper is as follows:
Section 2 introduces the related work.
Section 3 presents the details of the multimodal features used and the model architecture.
Section 4 describes the implementation details of the experiments and provides result analysis.
Finally, Section 5 summarizes our work.

\section{Related Work}
\label{sec:formatting}

\subsection{Video-based facial emotion analysis}
Video-based facial emotion analysis is a critical area of research within the field of affective computing, focusing on the automated recognition and interpretation of facial expressions from video data to infer the underlying emotional states of individuals, facial expressions play a vital role in understanding and analyzing emotions. Then a variety of pretrained models for facial expression recognition (FER)\cite{tian2011facial} or universal image analysis can be employed to extract frame-level visual features.  These include ResNet-Affecttet\cite{7780459, mollahosseini2017affectnet}, MANet-RAFDB\cite{li2017reliable, zhao2021learning}, AUs\cite{ekman1978facial}, FaceNet\cite{serengil2020lightface}, ViT\cite{caron2021emerging}, among others.  Notably, AUs provides an interpretable approach by considering the activation of specific facial muscles to encode facial expressions.  ResNet employs skip connections based on identity mappings to facilitate deep neural network training, while MANet combines a global multi-scale module and a local attention module for capturing both local and global information in facial emotion recognition.  As for Affectnet datasets, The dataset is a large-scale dataset widely used in research on emotion recognition. It consists of facial images sourced from the internet, annotated with seven emotion categories: anger, disgust, fear, happiness, sadness, surprise, and neutral. In summary, the AffectNet dataset's extensive size, containing hundreds of thousands of images, makes it a significant resource for studying emotion recognition and sentiment analysis.

\subsection{Audio-based emotion analysis}
In the realm of audio-based emotion analysis, feature extraction plays a crucial role in capturing discriminative information from acoustic signals to characterize emotional states accurately. Over the years, researchers have explored various feature extraction techniques to represent the complex dynamics of human emotions present in audio recordings. Earlier studies often relied on traditional manual feature extraction methods, such as Mel-frequency cepstral coefficients (MFCC) and extended Geneva Minimalistic Acoustic Parameter Set (eGeMAPS)\cite{eyben2015geneva}.  MFCC features capture the spectral characteristics of audio signals by transforming the power spectrum of the signal into the mel frequency domain and applying the discrete cosine transform.  MFCC provides a compact representation of the audio signal, capturing crucial spectral information related to mood. The feature can be extracted using the opensmil\cite{eyben2010opensmile} toolkit. On the other hand, eGeMAPS is specifically designed to capture a wide range of acoustic properties from speech signals, encompassing various acoustic parameters that cover spectrum, prosody, and speech quality characteristics.  eGeMAPS has emerged as a popular choice for speech-based emotion recognition and feature extraction in other related fields\cite{vlasenko2021fusion}. However, with the advent of deep learning, pre-trained models have gained widespread adoption in speech feature extraction tasks.  DeepSpectrum\cite{amiriparian2017snore}, for instance, leverages pre-trained convolutional neural networks originally designed for image recognition to extract acoustic features.  Its effectiveness has been demonstrated in various speech and audio recognition tasks.Furthermore, large models based on BERT\cite{devlin2018bert}, Wav2Vec\cite{schneider2019wav2vec} and Wav2Vec2.0\cite{baevski2020wav2vec} have also been applied to speech feature extraction, showcasing the potential of deep learning approaches in capturing complex patterns and dependencies in audio data.

\section{Approach}

\begin{figure*}
    \centering
    \includegraphics{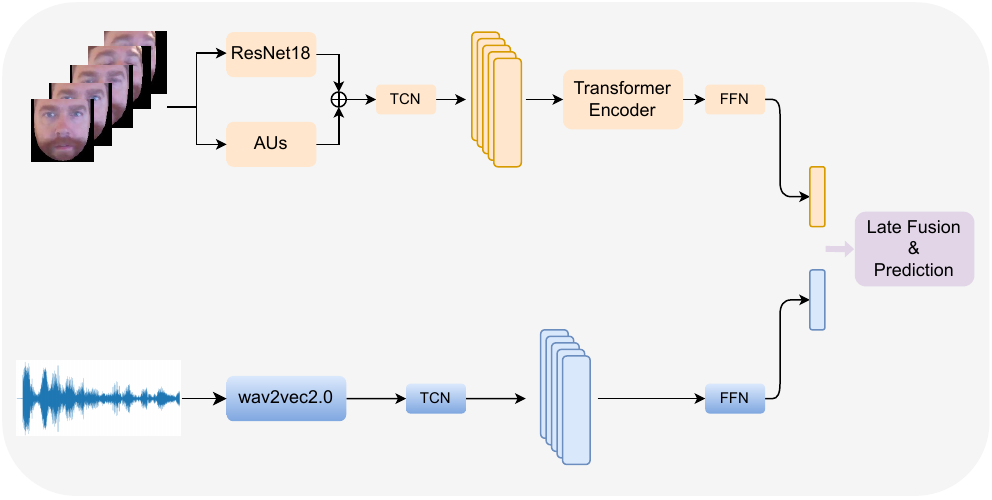}
    \caption{Model architecture}
    \label{fig:1}
\end{figure*}

In this section, we describe our method in detail, the architecture of the model is shown in Fig.\ref{fig:1} 
\begin{figure}
    \centering
    \includegraphics{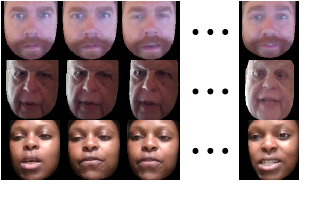}
    \caption{Preprocessed consecutive facial images}
    \label{fig:2}
\end{figure}

\subsection{Pre-processing}
For the EMI  estimation task involving video streams, the primary objective is to detect changes in facial expressions.  Therefore, preprocessing of the video data is of paramount importance.
We observed varying frame rates in the original videos, resulting in notable discrepancies in the number of extracted video frames. Consequently, we initially utilize the ffmpeg\cite{tomar2006converting} tool to standardize the frame rate of each video sample to 30 fps. For each video sample, we begin by using OpenCV\cite{deepak2021approach} to read the video as a series of consecutive frames. Subsequently, we conduct face detection on these frames, proceeding frame by frame. Frames containing detected faces are retained, while those without faces are discarded. For face detection and facial key point localization, we employ the Multi-task Cascaded Convolutional Networks (MTCNN)\cite{zhang2016joint} algorithm. Following face detection, we utilize the coordinates of facial key points provided by MTCNN to crop appropriate face regions from the images. These face regions are resized to 112x112 pixels and saved in the corresponding folders based on their sequence order, in preparation for subsequent feature extraction. The consecutive face images after preprocessing are shown in the Fig.\ref{fig:2}.

\subsection{Multimodal feature extraction}
\subsubsection{Videos features}
In the preprocessing stage, we have extracted continuous face images from the video, allowing us to directly extract facial expression features from these images.

\textbf{Vision Transformer (ViT)}: As an alternative vision-based approach, we utilize a DINO-trained ViT\cite{caron2021emerging}, which has undergone pre-training on the ImageNet-1K dataset using the Label-free Self-distillation (DINO) method. This model has demonstrated effectiveness across a range of image-based tasks, including facial expression emotion recognition \cite{chaudhari2022vitfer}. Upon processing the extracted facial images, the model generates a 384-dimensional embedding for each image. No additional pre-training or fine-tuning is performed on the model.

\textbf{ResNet18}: The Convolutional Neural Network (CNN) ResNet18, introduced by \cite{7780459}, is renowned for its exceptional ability to extract features from images.  In order to enhance its performance on facial datasets, a ResNet18 network pre-trained on AffectNet\cite{mollahosseini2017affectnet} is utilized to extract global spatial features from facial images.  The features before the final fully connected layer are averaged to obtain a 512-dimensional feature vector.  This improvement further enhances the network's capability to accurately extract relevant facial features.

\textbf{AUs}: The Facial Action Coding System (FACS)\cite{ekman1978facial} is a comprehensive method for objectively coding facial expressions.  In FACS, Action Units (AUs) correspond to specific facial muscles.  Each AU has two dimensions: the first dimension indicates detection, with 0 indicating absence and 1 indicating presence.  The second dimension represents intensity, ranging from 0 to 1.  We utilize OpenFace2.0\cite{krause2020automatic} to extract the detection and intensity of 17 AUs relevant to facial expressions.  Ultimately, each facial image receives a 34-dimensional feature embedding.

\subsubsection{Audios features}
Before extracting the audio features, we normalized all audio files to -3 dB and converted them to a format of 16 kHz, 16-bit mono.

\textbf{ Wav2Vec2.0}: Self-supervised pre-trained Transformer models have garnered considerable attention in the field of computer audition. A prominent example of such a foundational model is Wav2Vec2.0 \cite{baevski2020wav2vec}, which is frequently employed for Speech Emotion Recognition (SER) \cite{morais2022speech}. Given that all subchallenges are emotion-related, we leverage Wav2Vec2.0, specifically a large version fine-tuned on the MSP-Podcast\cite{lotfian2017building} dataset, for continuous emotion recognition. We derive audio signal features by averaging the representations from the final layer of the model, resulting in a 768-dimensional embedding.

\subsection{Temporal Encoder}
For the temporal encoder input, visual features are dual-channel, while acoustic features are single-channel.For the visual input features $X_v\in \mathbb{R} ^{{T} \times d_v}$, where $T$ represents the temporal dimension, while $d_v$ represents the spatial dimension.  The calculation formula is as follows:
\begin{equation}
    X_v = \text{Concate}(ResNet18(x), AUs(x))
\end{equation}
where $x\in \mathbb{R} ^{{T} \times H \times W \times 3}$ represents the input sequence of facial images, $T = 300$ represents the number of images, where H and W are both 112. "Concate" means to concatenate the two outputs along the spatial dimension.Similarly, acoustic features $X_a\in \mathbb{R} ^{{T} \times d_a}$ represent the feature sequence obtained from the Wav2Vec2.0 model.

Then, the visual features $X_v$ and acoustic features $X_a$ are fed into a Temporal Convolutional Network (TCN)\cite{chen2020transformer} based on one-dimensional causal convolution to gather local temporal context.TCN utilize dilated causal convolutions to capture temporal dependencies over long sequences efficiently. The architecture of TCN typically consists of multiple layers of dilated convolutional filters, followed by activation functions and possibly other layers such as pooling or normalization. The expression for a single layer of a TCN can be represented as follows:
\begin{equation}
    y_t = Relu(W \ast x_{t-d} + b)
\end{equation}
Where $y_t$ is the output at time step $t$. $ x_{t-d}$ represents the input sequence, with $d$ denoting the dilation factor.The dilation factor $d$ determines the receptive field of the convolutional filter and controls how many past time steps the filter can consider. $W$ is the learnable convolutional filter. $b$  is the bias term. $Relu$ represents the ReLU activation function. After passing through the TCN network, both the visual and acoustic features have the same spatial dimension $d_{model}$. In addition, behind the visual branch, a Transformer\cite{vaswani2017attention} Encoder is cascaded to interact with and integrate different parts of the input sequence, thereby extracting rich feature representations and capturing long-range dependencies in the sequence without introducing recursive structures.

\subsection{FFN}
Feedforward Neural Network (FFN) consists of two fully connected layers and a non-linear activation function to achieve non-linear mapping of input features. The input undergoes linear transformation and non-linear activation (ReLU) through the first fully connected layer, followed by linear transformation through the second fully connected layer to obtain the final output. Then the visual and acoustic output can be expressed as:
\begin{align}
    F_v = ReLU(X_vW_1 + b_1)W_2 + b_2 \\ 
    F_a = ReLU(X_aW_1 + b_1)W_2 + b_2
\end{align}
where \( W_1 \) and \( b_1 \) are the weight matrix and bias vector of the first fully connected layer, while \( W_2 \) and \( b_2 \) are those of the second fully connected layer. $F_v\in \mathbb{R}^6$ and $F_a\in \mathbb{R}^6$ respectively represent the outputs of the visual branch and the acoustic branch, both with a dimensionality of 6.

\subsection{Late Fusion and Prediction}
Our late fusion approach involves training the visual and acoustic models separately to make individual predictions, and then averaging the audiovisual results to obtain a final prediction from both modalities. The formula is as follows:
\begin{equation}
    F_{va} = average(F_v, F_a)
\end{equation}
where $F_{va}$ represents the final bimodal prediction result.

\subsection{Optimisation objective}
In this work, we utilize the mean square error (MSE) loss function in our training procedure. Let $y=[y_{1}, \cdots, y_{6}]$ and $\hat{y}=[\hat{y}_{1}, \cdots, \hat{y}_{6}]$ be the true emotional reaction intensity and the prediction, respectively, then the loss $\mathcal{L}$ can be defined as:
\begin{align}
    \mathcal{L} &= \text{MSE}(y, \hat{y})\\
   & =  \frac{1}{n} \sum_{i=1}^{n} (y_i - \hat{y}_i)^2 
\end{align}
where \( n \) represents the number of samples, \( y_i \) and \( \hat{y}_i \) denote the true label and predicted value of the \( i \)th sample, respectively.
\section{Experiments}

\subsection{Dataset}
In this work, we employed the multimodal Hume-Vidmimic2\cite{kollias20246th} dataset, which comprises over 15,000 videos featuring 557 participants and spanning more than 30 hours of audiovisual content.  Within this dataset, each participant was instructed to mimic a "seed" video depicting a person expressing a particular emotion.  Following the mimicry task, participants were prompted to assess the emotional intensity of the resulting video by selecting from a predefined set of emotional categories.  The dataset is speaker-independent and partitioned into training, validation, and test sets.  Table \ref{tab:dataset} presents the dataset statistics for each partition.

\begin{table}[htbp]
  \centering
    \begin{tabular}{ccc}
    \toprule
    Partition & Duration & Samples \\
    \midrule
    Train & 15:07:03 & 8072 \\
    Validation & 9:12:02 & 4588 \\
    Test  & 9:04:05 & 4586 \\
    \midrule
       $\sum$   & 33:23:10 & 17246 \\
    \bottomrule
    \end{tabular}%
    \caption{Hume-Vidmimic2 partition statistics.}
  \label{tab:dataset}%
\end{table}%

\subsection{Implement Details}

\textbf{Evaluation metric} Average Pearson’s Correlations Coefficient ($\rho$) is the metric used in intensity estimation, which is a measure of linear correlation between predicted emotional reaction intensity and target, then the metric can be defined as follows:
\begin{equation}
\rho = \sum_{i=1}^{6} \frac{\rho_{i}}{6}
\end{equation}
where $\rho_{i}(i \in \{1,2\cdots,6\})$ for 6 emotions, respectively, and is defined as:
\begin{equation}
\rho_{i} = \frac{\text{cov}(y_{i}, \hat{y_{i}})}{\sqrt{\text{var}(y_{i})\text{var}(\hat{y_{i}})}}
\end{equation}
where $\text{cov}(y_{i}, \hat{y_{i}})$ is the covariance between the predicted value and the target, $\text{var}(y_{i})$ and $\text{var}(\hat{y_{i}})$ are variance respectively.

\textbf{Training settings} The training process is optimized using the Adam optimizer \cite{kingma2014adam}. All experiments were conducted on an NVIDIA RTX 3090 GPU with PyTorch, using an initial learning rate of $3e^{-5}$ and a batch size of 128. Additionally, if the validation set metric does not improve for 10 epochs, the learning rate is halved. The visual branch encoder has a dimension of 546 with 2 encoder blocks and 4 multi-heads. In the temporal encoder, the 1-dimensional convolution kernel size is 3, there are 5 convolution layers, and the feature dimension in attention is 128.

\subsection{Results}

\textbf{Unimodal Results}. For the EMI estimation challenge, we initially assessed the effectiveness of our unimodal features (video, audio) on the validation set and compared them with the officially provided features. The experimental results are presented in Table \ref{tab:unimodal}. It is evident from Table \ref{tab:unimodal} that combining our extracted features with the official ones led to performance improvements over the baseline on the validation set. Concerning the model, retraining the official features with our model resulted in notable enhancements in the visual and audio modalities, with increases of 1$\%$ and 4.08$\%$, respectively. Furthermore, our experiments with ResNet18 and AUs for visual feature extraction yielded significant improvements of 0.1236 and 0.1352, respectively. Finally, by integrating the visual dual-channel features ( ResNet18, AUs), we achieved a visual result of 0.1479, thus confirming the efficacy of our extracted multi-channel features.

\begin{table}[htbp]
  \centering
    \begin{tabular}{ccc}
    \toprule
    Features & Modality & Mean $\rho$ \\
    \midrule
    ViT(baseline)\cite{kollias20246th} & $V$     & 0.09 \\
    ViT(ours) & $V$     & 0.10 \\
    ResNet18 & $V$     & 0.1236 \\
    AUs   &  $V$     & 0.1352 \\
    ResNet18+AUs & $V$     & \textbf{0.1479} \\
    \midrule
    Wav2Vec2.0(baseline)\cite{kollias20246th} & $A$     & 0.24 \\
    Wav2Vec2.0(ours) & $A$     & \textbf{0.2808} \\
    \bottomrule
    \end{tabular}%
    \caption{The unimodal results on validation set of the EMI Estimation Challenge. We report the Pearson correlation coefficient ($\rho$) for the average of 6 emotion targets. Where $V$ represents the visual modality, and $A$ represents the acoustic modality.}
  \label{tab:unimodal}%
\end{table}%

\textbf{Multimodal Results.} After obtaining the unimodal prediction results, we utilized a late fusion strategy to obtain audiovisual prediction results. The experimental results are presented in Table \ref{tab:multimodal}. From Table \ref{tab:multimodal}, it is evident that our multimodal results outperform the official ones. Ultimately, by averaging the visual dual-channel features (ResNet18, AUs) and the acoustic single-channel feature (Wav2Vec2.0), we achieved a performance of 0.3288 on the validation set, representing a 7.88 percentage point improvement over the baseline.

\begin{table}[htbp]
  \centering
    \begin{tabular}{ccc}
    \toprule
    Modality & Features & Late Fusion \\
    \midrule
             & ViT+Wav2Vec2.0(baseline)\cite{kollias20246th} & 0.25 \\
            & ViT+Wav2Vec2.0(ours) & 0.2835 \\
     $V$+$A$     & ResNet18+Wav2Vec2.0 & 0.2983 \\
              & AUs+Wav2Vec2.0 & 0.3026 \\
            & ResNet18+AUs+Wav2Vec2.0 & \textbf{0.3288} \\
    \bottomrule
    \end{tabular}%
    \caption{Multimodal results of late fusion for mean $\rho$ on the validation set.}
  \label{tab:multimodal}%
\end{table}%
\section{Conclusion}
In this paper, we present the solution to the Emotional Mimicry Intensity (EMI) Estimation challenge, which is part of 6th Affective Behavior Analysis in-the-wild (ABAW) Competition. We obtained effective feature representations by extracting visual dual-channel features (ResNet18, AUs) and acoustic single-channel feature (Wav2Vec2.0). Subsequently, based on a late fusion strategy, we fused the audiovisual results. Experimental validation demonstrated the effectiveness of our proposed approach, achieving an average $\rho$ of 0.3288 on the validation set.
{
    \small
    \bibliographystyle{ieeenat_fullname}
    \bibliography{main}
}


\end{document}